# Energy-leaky modes in partially measured scattering matrices of disordered media

Hyeonseung Yu[1], KyeoReh Lee[1], and YongKeun Park[1,*]

[1]Department of Physics, Korea Advanced Institute of Science and Technology, Daejeon 34141, South Korea

*We investigate energy-leaky modes induced by incomplete measurements of the scattering matrices of complex media. Due to the limited numerical apertures of an optical system, it is experimentally challenging to access theoretically predicted perfect transmission channels in the diffusive regime. By conducting numerical simulations on scattering matrices, we demonstrate that leaky modes contributed from uncollected transmission in the TM provides an energy transmission that is more enhanced than that predicted by measurement. On the other hand, a leaky mode originating from the uncollected reflection in the partial measurement of a RM strongly suppresses the energy transmission through a zero-reflection channel, restricting the transmission enhancement to no more than a 5-fold enhancement in limited optical systems. Our study provides useful insights into the effective control of energy delivery through scattering media and its ultimate limitation in practical schemes.*

Light transport through complex media is a fundamentally important physical problem and also has implications in various applications. Recently, the active control of light transport through complex media by shaping an impinging wavefront has been demonstrated, and gained significant interest [1-4].

Such phenomena, governed by the interference of multiply scattered waves, can be well formulated with a scattering matrix (SM). A SM describes a linear relationship between input and output complex fields. A SM consists of two transmission matrices (TM) and two reflection matrices (RM), and considers illuminations from both sides. Because a SM contains the full optical information about light transport through a turbid medium, it has been exploited in various studies, including image reconstruction through a scattering layer [5] a multimode fiber [6], a spectrometer [7], biophotonics [4,8-11] [12], plasmonic control [13], near-field control [14,15], lasers [16], photoacoustics [17], and photovoltaic systems [18].

One of the most important questions in light transport though turbid media concerns energy transport. Theoretically, the existence of perfect transmission channels, also known as open channels, have been predicted even in highly scattering media, by the random matrix theory [19,20] and particularly the DMPK equation [21,22]. The transmission eigenvalues of ideal TMs or RMs, manifesting energy transmittance, follow a bimodal distribution when wave propagation is in the diffusive regime, which indicates the existence of perfect transmission channels regardless of sample thicknesses.

Unfortunately, attempts to enhance energy delivery through turbid media via open channels have not been fully explored in experimental conditions. Recently, several works have demonstrated enhancements of energy delivery [23,24]. However, the demonstrated enhancements were far below the theoretically expected ones, mainly due to the limited number of optical modes for measurements and controls. More recently, it was shown that the limited numerical apertures (NAs) of lenses in optical systems prevent them from accessing information about open channels, and thus enhanced energy delivery through turbid media will be significantly limited [25,26]. Although the accessible information in experimental conditions has been well described, uncollected information which is beyond the capability of a measurement system, and its effect on energy transmission, have never been previously considered.

Recently, several methods have been developed to enhance energy delivery via reflection measurements [27,28]. However, transmission was restricted to only about 3-fold enhancements in these studies. In addition, the ultimate limit of the enhancement was not clarified, particularly in relation to practical experimental conditions. Importantly, controlling energy transmission by monitoring reflected fields also has potential for biomedical applications because it does not require an invasive scheme.

In this letter, we investigate energy leaky modes which originate from partial measurements of a SM, using numerical simulations. When a TM or RM of a medium is completely measured, incident energy can be fully delivered to the output side of the medium by employing open channels. We show that when a TM or RM is incompletely measured due to the limited NA, significant energy will be coupled out beyond the NA of the optical experimental systems. These energy leaky modes provide enhanced energy delivery in the transmission geometry, and manifest strong suppression of energy transmission in the reflection geometry. The presented concept of the energy leaky modes can clearly explain the deviation between experimental results and the theoretical expectation, and also provide the practical limitations of energy transmission in actively controlled light transport through turbid media.

The leaky modes resulting the partial measurement of a SM are illustrated in Fig. 1. Before proceeding, we define a measure $f$ to be the fraction of experimentally accessible optical modes to the total optical modes of a scattering medium [26]. Here, $f$ is formulated as $f = (NA/n)^2$, where $NA$ and $n$ denote the NA of an objective lens and the refractive index of a surrounding medium, respectively. For simplicity and without losing generality, we also assume that the NAs of illumination and collection are the same.

In an ideal situation, $f = 1$ [Fig. 1(a)], a TM can be perfectly measured; all existing optical modes in turbid media can be fully accessed with their transmission response. Then, an incident wave field can be coupled into a perfect open channel, by using a spatial light modulator for example, and all input energy can be fully transmitted to the output side, i.e. transmittance $\tau = 1$.

When a TM is partially measured due to the limited NA, $f < 1$ [Fig. 1(b)], complete access to an ideal open channel is impossible. In this case, the maximal transport channel has transmittance smaller than unity, $\tau_{max} < 1$ [25,26]. Nonetheless, there exists transmitted but uncollected energy $T_{leaky,f}$, which is referred to as leaky modes. The leaky modes can contribute to the enhanced energy delivery depending on applications, even though it is not directly collected by the objective lens.

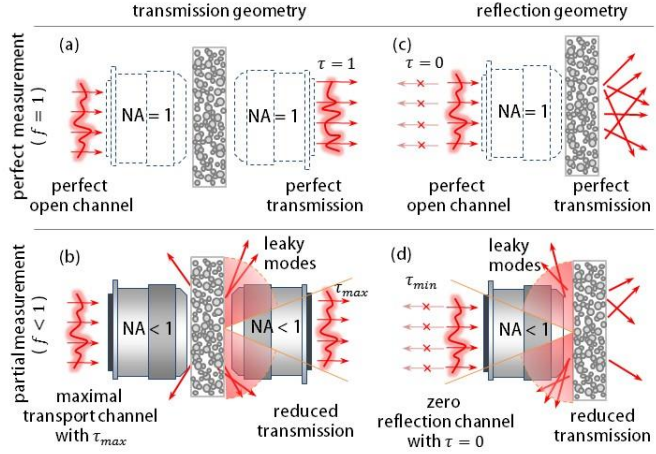

FIG. 1 (color online) Open channels and leaky modes in various experimental schemes. (a) Complete measurement of a TM allows access to perfect transmission, $\tau = 1$. (b) When a TM is partially measured, transmission through the maximal transport channel $\tau_{max}$ is significantly decreased, although the total transmitted energy through the turbid media $T_f$ is high, $\tau_{max} < T_f < 1$. This is because some energy is coupled out to leaky modes, even though it is not collected by the detection lens (the shaded area). (c) Similarly, a complete measurement of a RM enables the full transmission of incident energy through the perfect zero-reflection channel, i.e. $\tau = 0$. (d) Partial measurement of a RM produces significant light loss through leaky modes, and transmittance via the zero-reflection channel is severely suppressed.

Similarly, a perfect measurement of a RM enables the transfer of full incident energy through a scattering medium via a zero-reflection channel [Fig. 1(c)]. When light absorption is neglected, an open channel in the TM corresponds to a zero-reflection channel acquired from the RM. In the case of a partial measurement of a RM, $f < 1$, a zero-reflection channel still exists. However, the zero-reflection channel does not guarantee that the reflection beyond the NA of an objective lens vanishes. This uncollected reflection $R_f$ corresponds to a leaky mode in reflection geometry. The leaky modes of a RM suppress energy transmission through a zero-reflection channel.

To quantitatively analyze the effects of leaky modes on light transmission through turbid media, we numerically simulated SMs and investigated the energy transport via leaky modes. The set of SMs of scattering media in the diffusive regime is obtained based on an approach used in disordered metallic systems [29]. The simulation assumes a scattering medium as a multilayer of weakly scattering thin slices. By changing the number of slices, SMs with various optical thickness are obtained. In our simulation, the numbers of input and output modes were both set to be $n = 4,096$. The optical thickness is $L / l_s = 30$, where $L$ is the physical thickness of the scattering medium and $l_s$ is the mean free path of light scattering. A simulated SM is decomposed into two TMs and two RMs as follows:

$$S = \begin{bmatrix} r & t \\ t' & r' \end{bmatrix}. \qquad (1)$$

Here we consider only one-sided illumination onto the sample, so our scope is restricted to $r$ and $t$.

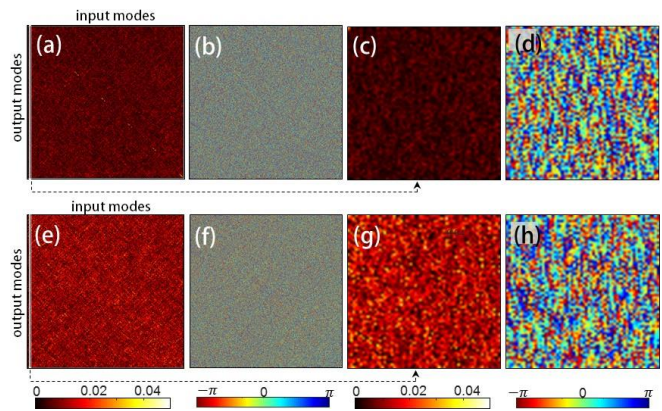

FIG. 2 (color online) (a) Amplitude and (b) phase of a simulated TM. (c) Amplitude and (d) phase maps corresponding to a representative input mode (white rectangular area). (e) Amplitude and (f) phase of a simulated RM. (b) Amplitude and (b) phase maps corresponding to a representative mode (white rectangular area).

Simulated TM and RM are shown in Fig. 2. The amplitude [Fig. 2(a)] and phase part [Fig. 3(a)] of the TM show uncorrelated random complex entities. The first column of the matrix, which corresponds to the output field response to a single input wave function, is represented in a 2-D spatial map with the amplitude [Fig. 2(c)] and the phase part [Fig. 2(d)]. Similarly, the amplitude [Fig. 2(e)] and phase part [Fig. 2(f)] of the RM and the representative 2-D field map in amplitude [Fig. 2(g)] and in phase [Fig. 2(h)] exhibit random properties. The mean reflectance of the RM is significantly greater than that of the TM, which is the signature of a highly diffusive regime.

Using the simulated SMs, we first studied the energy transport in the measurement of the TM. When measured with an optical system with $f$, only a fraction of an ideal TM $t$ can be accessed, and this measurable TM $t_f$ has $m$ input and output modes, where $m = fn$. In our numerical simulation, $t_f$ can be obtained by taking an $m \times m$ submatrix of the full TM $t$. Then, the eigenvalue decomposition is applied to $t_f^\dagger t_f$ in order to find the maximal transport channel,

$$t_f^\dagger t_f = U_f \Lambda_f V_f, \qquad (2)$$

where $V_f = [v_1\ v_1\ \ldots\ v_m]$ contains transmission eigenvectors.

Let $\tau_{max,f}$ and $v_{max,f}$ be the maximum eigenvalue and its corresponding eigenvector. It is noteworthy that this maximal transport eigenvector is not the same as the open channel acquired from the full matrix. In a real experiment, the actual input field vector is formulated as the normalized $n$-dim vector whose $m$ components are equal to $v_{max,f}$ and the remaining $n-m$ components are zero,

$$E_{max,f} = \frac{1}{\sqrt{|v_{max,f}|}}\left[v_{max,f}\ \ 0\ \ \ldots\ \ 0\right]. \qquad (3)$$

Then, we define the total transmitted energy $T_f$ as follows,

$$T_f = |tE_{max,f}|^2. \qquad (4)$$

Here we assumed that the total transmitted energy is measured with an ideal objective lens with $f = 1$, although the TM is partially measured. The mean transmittance $\langle T \rangle_f$ of the medium is defined as the transmitted energy for a plane wave illumination $I_f$.

$$I_f = \frac{1}{\sqrt{m}}[1\ \ \cdots\ \ 1\ \ 0\ \ \cdots\ \ 0],\text{ and} \qquad (5)$$

$$\langle T \rangle_f = |tI_f|^2. \qquad (6)$$

The numerically simulated energy transported by the maximal eigenchannel and leaky mode are shown in Fig. 3(a). The mean transmittance of the scattering medium is $0.0818 \pm 0.0085$ for the entire range of $f$, hence it has no dependence on $f$. When a sample is illuminated with the maximal transport channel, the delivered energy, which is equal to the maximum eigenvalue $\tau_{max,f}$, linearly decreases with $f$. This maximum eigenvalue is the quantity that is experimentally measured. Interestingly, the total transmitted energy $T_f$, including the energy transmitted by both the maximal transport channel and the leaky mode, is larger than the maximal eigenvalue. This indicates that the actual energy transport is more enhanced than one can expect from the measurement. For example, a predicted transmission is $0.5038 \pm 0.0293$ in a measurement with $f = 0.5$. However, the actual total transmittance is $0.6845 \pm 0.0152$, which is 35% enhanced over the predicted value.

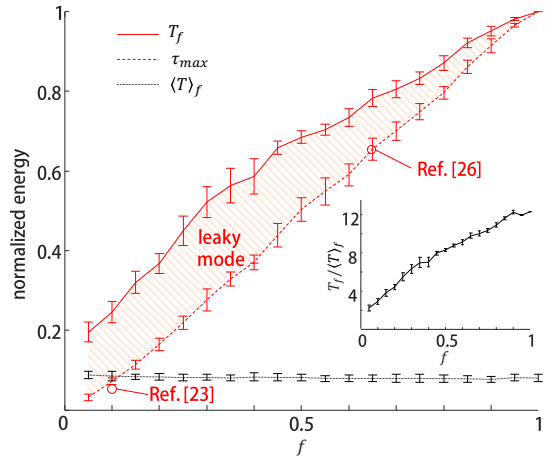

FIG. 3 (color online) Energy transmission through the maximal transport channel, $\tau_{max}$ (red solid line) and the total transmitted energy, $T_f$ (red dashed line) as a function of $f$. The deviations between $T_f$ and $\tau_{max}$ correspond to the leaky modes (orange shaded area). The gray solid line represents the mean transmittance $\langle T \rangle$. (*inset*) The enhancement factor, $T_f/\langle T_f \rangle$ as a function of $f$.

We also compared our simulated results with reported values from previous experimental reports. In an experiment on the maximal energy transport [23], the mean transmittance was about 0.079, taking into account the limited collection angle. This value is coincident with the mean transmittance 0.0818 of our simulated TM. In Ref. [23], the optical system had $f = 0.1$ and the maximal eigenvalue was approximately equal to

0.031. This value is compatible with our simulated result of $0.072 \pm 0.0086$, where the smaller eigenvalue in the experiment can be explained by the fact that the illumination NA was smaller than the collection NA. According to our simulated result, total transmittance of 0.246 would have been achieved rather than 0.031, if they had collected leaky modes as well as optical modes transmitted through the objective lens.

In our previous study [26], a TM was experimentally measured with $f = 0.65$ and the maximum eigenvalue was reported to be 0.65, which is consistent with our simulated values of $0.6546 \pm 0.0279$. If the maximal transport channel had been exploited, the total transmittance of $0.7827 \pm 0.0210$ could have been obtained, which provides significantly enhanced light transmission.

In experimental realizations of maximal transport channels, the practically important quantity is the enhancement of delivered energy over the mean transmittance. Since the mean transmittance is constant over $f$, the trend of the enhancement factor, defined as $T_f/\langle T_f \rangle$, follows the curve shape of the total transmitted energy (*inset*, Fig. 3). The enhancement factor linearly increases with $f$, reaching up to approximately 12. Because the maximal eigenvalue is a universal quantity [30], a higher enhancement factor can be obtained with thicker samples with smaller mean transmittance.

Next, we examined the role of leaky modes in a reflection geometry. Similar to the case of a TM, a zero-reflection channel is retrieved from an $m \times m$ partial RM $r_f$, where $m = f \cdot n$. To find an open channel of $r_f$, the eigenvalue decomposition is applied to $r_f^\dagger r_f$,

$$r_f^\dagger r_f = P_f \Lambda_f Q_f , \qquad (7)$$

where $Q_f = [q_1\ q_2\ \ldots\ q_m]$ contains reflection eigenvectors. For the minimum eigenvalue $\tau_{min,f}$ and the corresponding zero-reflection eigenvector $q_{min,f}$, the input field vector in an experiment is given as follows,

$$E_{min,f} = \frac{1}{\sqrt{|q_{min,f}|}} \begin{bmatrix} q_{min,f} & 0 & \ldots & 0 \end{bmatrix}. \qquad (8)$$

Then, the energy loss $R_f$ through the leaky mode and the total transmitted energy $T_f$ is calculated as,

$$R_f = |rE_{min,f}|^2 \qquad (9)$$

$$T_f = 1 - R_f - \tau_{min} . \qquad (10)$$

The mean transmittance of the media is defined in the same way as in the case of transmission geometry.

Compared to the transmission case, there are two distinctive characteristics in the leaky modes in a RM: (1) the zero-reflection channel always exists regardless of $f$ [31]. One can assume $\tau_{min}$ is zero, and (2) a leaky energy $R_f$ in a RM is defined as uncollected reflected light beyond the limit of NA, and thus the leaky mode suppresses the energy transmission, as seen in Eq. (10).

The total transmitted energy $T_f$ corresponding to the illumination via a zero-reflection channel as a function of $f$ is plotted in Fig. 4(a). Notably, when $f$ is slightly below unity, energy transmission via a zero-reflection mode decreases significantly. For example, the transmittance is only $0.2275 \pm 0.0401$ even with $f = 0.8$. Here $f = 0.8$ corresponds to an objective lens with $NA = 0.9$. Furthermore, the transmittance becomes smaller than 0.2 when $f < 0.7$. This result indicates that it is crucial to measure the full RM to exploit open channels in a reflection geometry. In our recent experimental work [31], a RM was measured with $f = 0.25$. Although the zero-reflection channel was observed in the acquired RM, the expected total transmitted energy would be only $0.113 \pm 0.0117$ according to the current result.

Similar to the TM case, the enhancement of the energy as a function of $f$ follows the same trends with the total transmitted energy in the reflection geometry [Fig. 4(b)]. The enhancement factors are less than 4 except when $f = 1$, suggesting the predicament of the experimental realization of exploiting open channels. This limited controllability is also found in recent experiments [27,28].

To compare the presented work with recent experimental work, we analyzed the expected transmission considering the used experimental setups. In Ref. [27], the illumination NA ($f_I = 0.2$) and the collection NA ($f_C = 1.0$) were different. Similarly, the setup in Ref. [28] had mismatched NAs ($f_I = 0.64$, $f_C = 1.0$). Due to the assumption of the symmetry between illumination and collection NAs in our simulation, we took the geometric mean $f_{mean} = \sqrt{f_I f_C}$. Then the experimental conditions in Refs. [27] and [28] correspond to $f_{mean} = 0.44$ and $f_{mean} = 0.8$, respectively. The experimentally observed enhancement factors are indicated in Fig. 4(c) and both values are well matched with our numerical simulation.

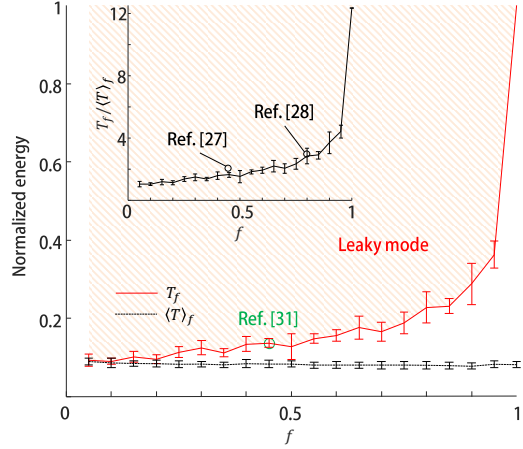

FIG. 4 (color online) Energy transmission through the zero-reflection channels, $T_f$ (red line) with significant energy loss due to the leaky modes (orange shaded area). In reflection geometry, $\tau_{max} = 1$ regardless of a fraction number $f$. Gray solid line represents the mean transmittance, $\langle T_f \rangle$. (inset) The enhancement factor, $T_f/\langle T_f \rangle$ as a function of $f$.

In order to study the effect of the scattering strength of a scattering medium, we finally investigated the dependence of the leaky modes on optical thickness. We simulated scattering media with three different thicknesses: $L/l_s = 30, 50,$ and $70$. Governed by the Ohm's law in the diffusive limit [32], the mean transmittance of scattering media is inversly propotional to the sample thickness. The total transmitted energy for maximal energy transport channels in the TM is shown as a function of $f$ [Fig. 5(a)]. It is observed that the energy transmission decreases as the optical thickness increases. This result indicates that the energy transport through leaky modes is governed by the mean transmittance of the scattering medium. In the RM, the energy transmission is more strongly prohibited as the sample thickness increases [Fig. 5(b)].

The enhancements of energy tranmission show slightly different characteristics in the TM and RM [Fig. 5(c)]. In the TM case, the enhancement was larger for thicker samples. This is due to the fact that the energy transmission through the maximal transport channel is independent of the optical thickness [30], while the mean tranmittance linearly decreases as the sample becomes thicker. In contrast, the enhancement in the RM remains almost constant over the entire range of $f$ except near unity. This result implies that the leaky mode in the RM is strongly correlated with the mean transmittance so that the energy transmission through a zero-reflection channel is dominated by the mean transmittance except when $f = 1$.

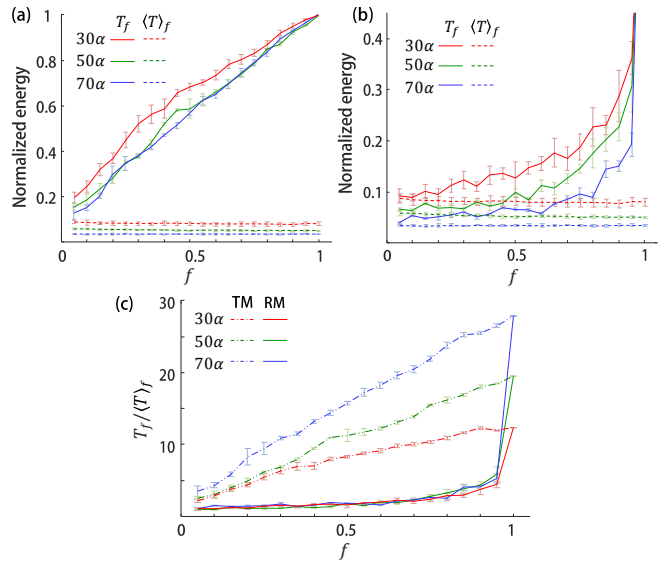

FIG. 5 (color online) (a) Total transmitted energy thorough maximal transport channels in TMs (solid lines) and the mean transmittance (dashed lines) with various optical thicknesses. (b) Total transmitted energy through zero-reflection channels in RMs (solid lines) and the mean transmittance (dashed lines) with various optical thicknesses (c) The enhancement factor in TMs (dashed dot lines) and RMs (solid lines).

In summary, we present that partial measurements of SMs produce energy leaky modes, which contibute to the enhanced delivery in the TM and severe prohibition of energy transfer in the RM. In particular, efficient energy transmission is intrinsically restricted in a partially measured RM because the minimal reflection channel is strongly mismatched with the genuine open channel of the scattering media. The numerically simulated results showed that it is extremely difficult to achieve perfect transmission in the reflection geometry; an approximate factor of 5 is the maximum achievable limit of transmission enhancement in practical optical systems.

From a practical point of view, this work provides useful insight into the optimal strategy to realize efficient energy transfer through highly scattering media. In order to address various experimental schemes, further studies are necessary, including the relation between the imperfect coupling to open channels and spatial profiles inside a medium [33], effects of different geometry factors of scattering medium [34] or near-field propagation [14,15,35]. Our work can be extended to other optical regimes, such as Anderson localization [36], absorbing [25,37,38] or amplifying [39] scattering media. We also expect that more consideration of various optical degrees, for example the time-resolved RMs [40], will provide a more efficient way of energy transfer through scattering media.

This work was supported by KAIST, and the National Research Foundation of Korea (2015R1A3A2066550, 2014K1A3A1A09063027, 2012-M3C1A1-048860, 2014M3C1A3052537) and Innopolis foundation (A2015DD126).

* yk.park@kaist.ac.kr